\documentclass{moriond}

\usepackage{multirow}
\def\Journal#1#2#3#4{{#1} {\bf #2}, #3 (#4)}

\def\be{\begin{equation}}
\def\ee{\end{equation}}
\def\bea{\begin{eqnarray}}
\def\eea{\end{eqnarray}}

\newcommand{\Photo}{\includegraphics[height=35mm]{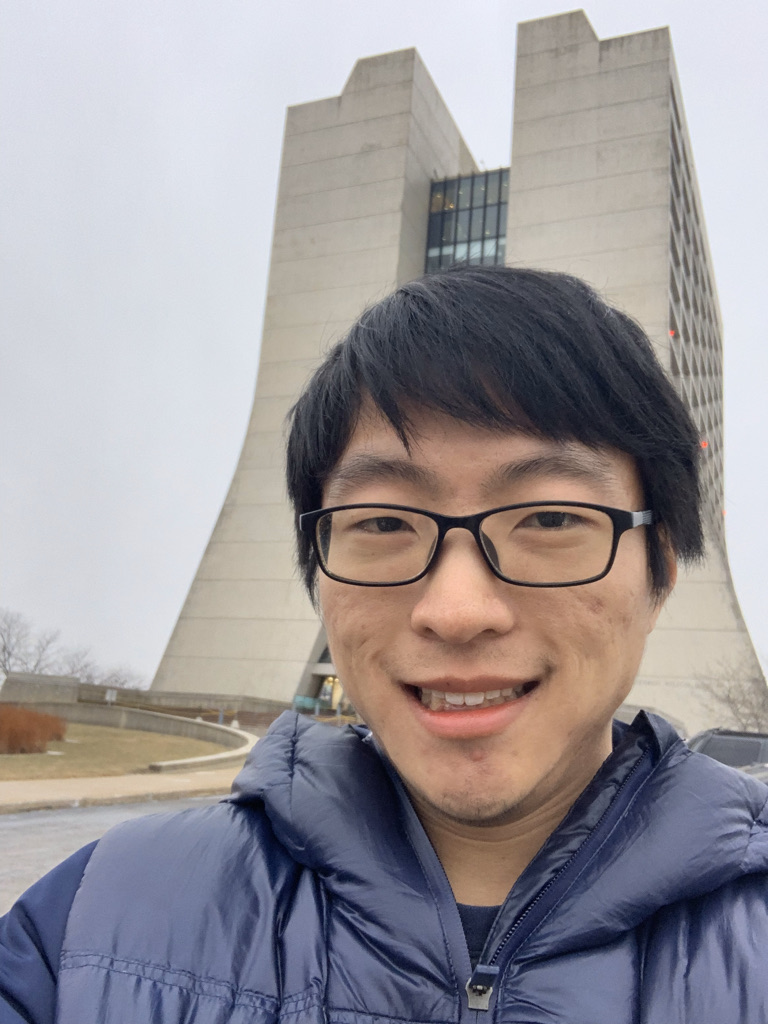}}

\begin{document}
\vspace*{4cm}
\title{Search for charged-lepton flavor violation in the production and decay of top quarks using trilepton final states}

\author{Jingyan Li on behalf of the CMS Collaboration}

\address{Department of Physics, Northeastern University,\\
360 Huntington Ave, Boston, United States}

\maketitle\abstracts{
This document describes a search for charged-lepton flavor violation (CLFV) in the production and decay of top quarks using 138 fb$^{-1}$ of data collected by the CMS experiment at a center-of-mass energy of 13 TeV. Events are selected for analysis if they contain an opposite-sign electron-muon pair, a third charged lepton (electron or muon), at least one jet, and at most one jet associated with a bottom quark. The analysis utilizes boosted decision trees to separate background processes from a possible signal. The data were found to be consistent with the standard model expectation. Exclusion limits were placed on different CLFV interactions, constituting the most stringent limits to date on these processes. }

\section{Introduction}

Charged-lepton flavor violation (CLFV) is forbidden in the standard model (SM) with massless neutrinos. However, the connection \cite{Theory1} between CLFV and flavor anomalies in decays of B mesons \cite{LHCbExp,LHCb} provides hints of observable CLFV effects in the top quark sector at the TeV scale. The CLFV signatures involving top quarks have been studied phenomenologically \cite{Theory2} and experimentally by both the ATLAS \cite{ATLASExp,ATLAS} and CMS \cite{CMSExp,TOP-19-006} Collaborations, yielding strong constraints on CLFV branching fractions.

\begin{figure}[h]
 \begin{center}
  \begin{tabular}{ccc}
   \includegraphics[width=0.325\textwidth]{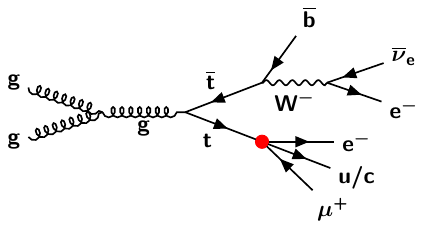}&
   \includegraphics[width=0.325\textwidth]{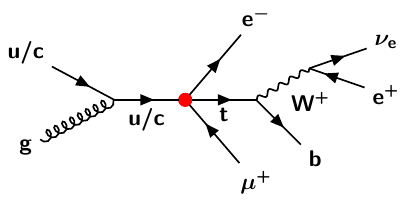}&
   \includegraphics[width=0.325\textwidth]{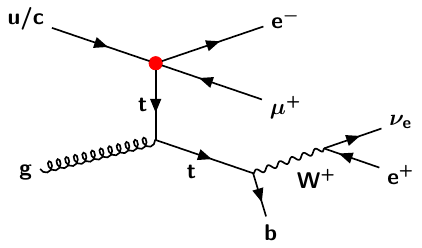}
  \end{tabular}
 \caption{Representative Feynman diagrams for top decay (left) and production (center and right) CLFV processes considered by this analysis.}
 \label{fig:Feynman_Diagram}
 \end{center}
\end{figure}

In this document, we report a model-independent CLFV search \cite{TOP-22-005} targeting both top production and decay CLFV signals in trilepton final states; either electrons or muons are considered. The search utilizes the proton-proton collision data collected by the CMS experiment at the LHC in 2016--2018 at a center-of-mass energy of 13 TeV, corresponding to an integrated luminosity of 138 fb$^{-1}$. The CLFV signals are parameterized with Dimension-6 Effective Field Theory (EFT) operators. Representative Feynman diagrams are shown in Figure \ref{fig:Feynman_Diagram}.

\section{Event selection and background modeling}

The target final state of this analysis is characterized by three charged isolated leptons originating from the CLFV interaction or decays of electroweak bosons. These leptons, referred to as ``prompt" leptons, differ kinematically from ``nonprompt" leptons that originate from the decays of hadrons, or from photon conversions. The signal region (SR) requires events to contain three leptons, selected with tight identification criteria, at least one jet and no more than one b-tagged jet. Events with an opposite-sign, same-flavor (OSSF) lepton pair with an invariant mass close to the Z boson mass are removed from the SR as well.

The SM backgrounds that produce at least 3 prompt leptons are referred to as ``prompt backgrounds". These backgrounds are modeled with MC simulation. The largest contribution to the prompt backgrounds comes from the production of WZ bosons. The modeling of WZ production is validated in dedicated control regions that feature at least one OSSF lepton pair. 

\begin{figure}[h]
 \begin{center}
  \begin{tabular}{ccc}
   \includegraphics[width=0.325\textwidth]{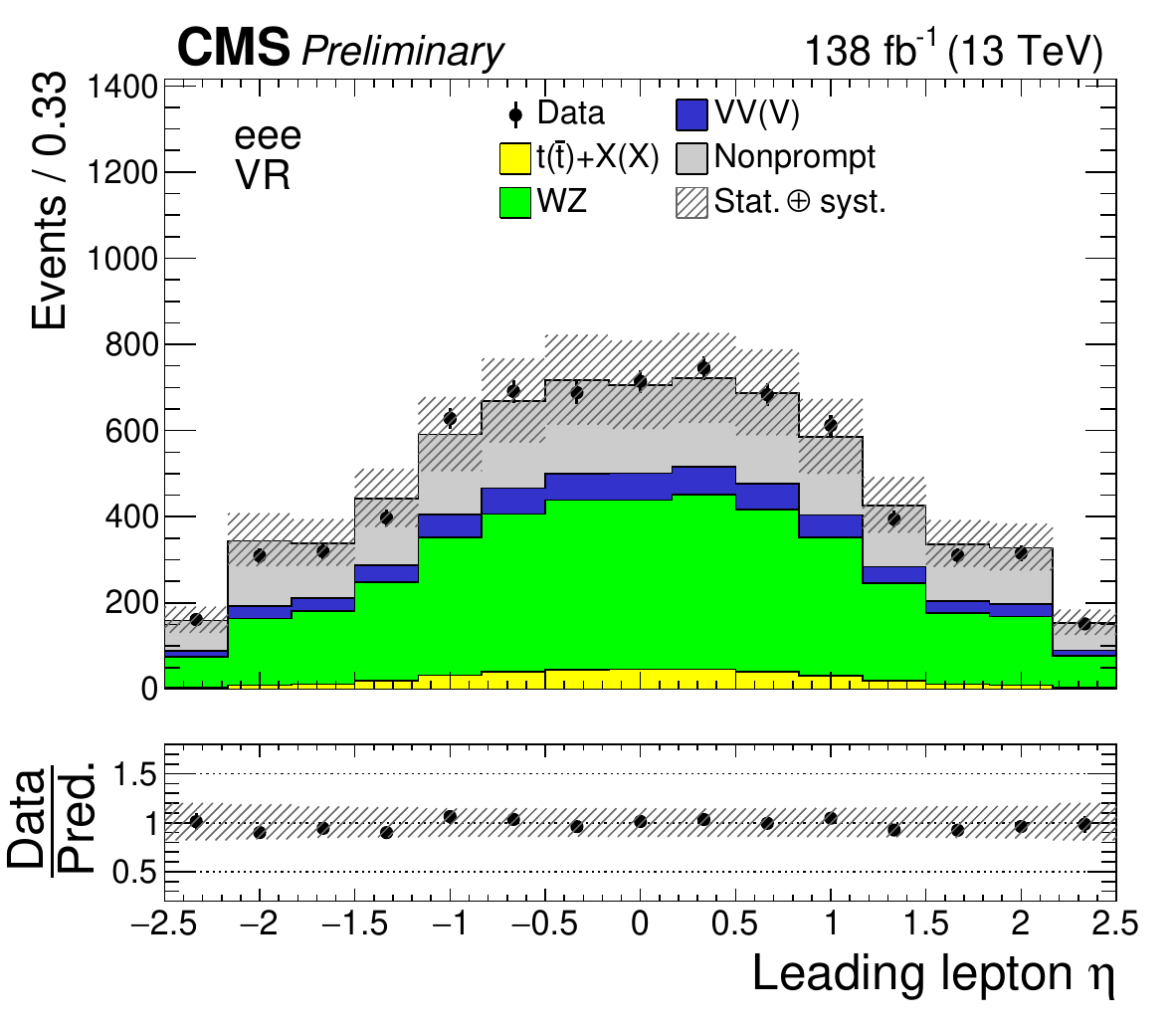}&
   \includegraphics[width=0.325\textwidth]{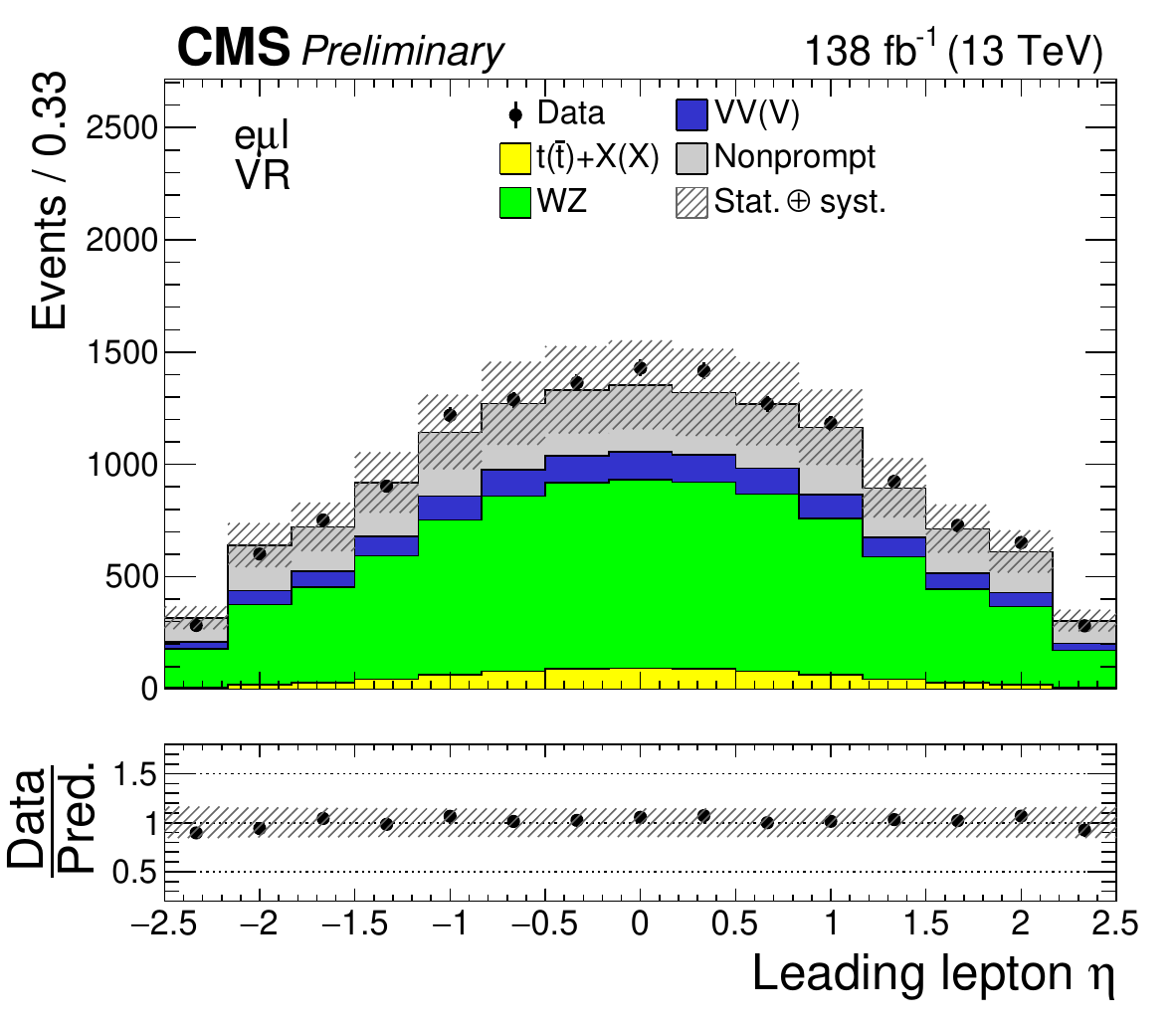}&
   \includegraphics[width=0.325\textwidth]{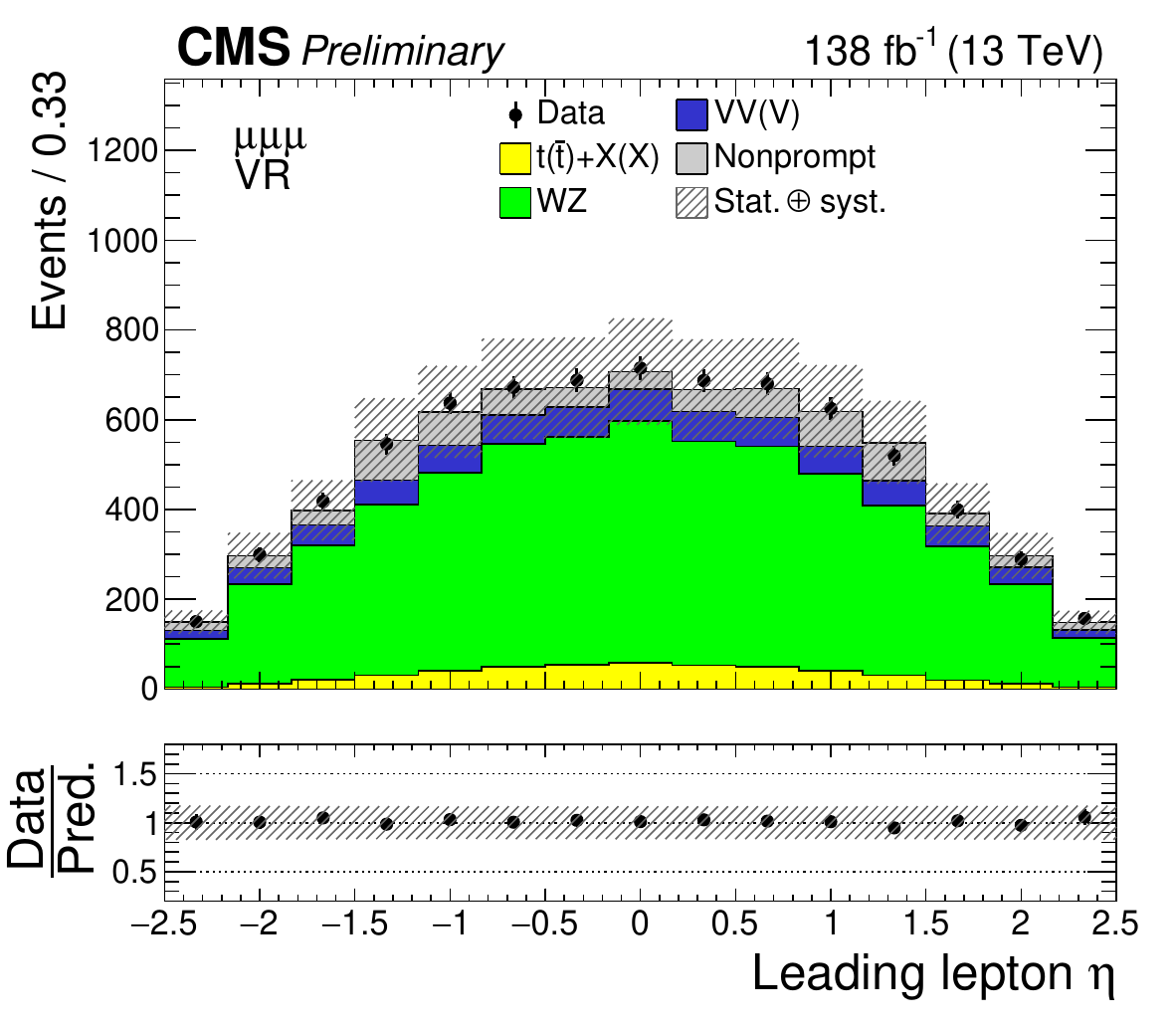}
  \end{tabular}
 \caption{The pseudorapidity distributions of the leading-$p_{T}$ lepton in VRs. Events are selected with $eee$ (left), $e\mu\ell$ (center), $\mu\mu\mu$ (right).}
 \label{fig:VR}
 \end{center}
\end{figure}

Despite the stringent lepton selection criteria, nonprompt leptons can still enter the event selection contributing to the ``nonprompt backgrounds". Drell-Yan and $t\bar{t}$ production are the leading contributors to the nonprompt backgrounds. A robust modeling of the nonprompt backgrounds is difficult to achieve through MC simulation, and therefore, a data-driven technique called ``the matrix method" \cite{Matrix} is used. The validity of this method is evaluated in validation regions (VRs), where good agreement between measurement and prediction is observed, as is shown in Figure \ref{fig:VR}.

\section{Signal extraction and statistical analysis}

\begin{figure}[h]
 \begin{center}
  \begin{tabular}{cc}
   \includegraphics[width=0.45\textwidth]{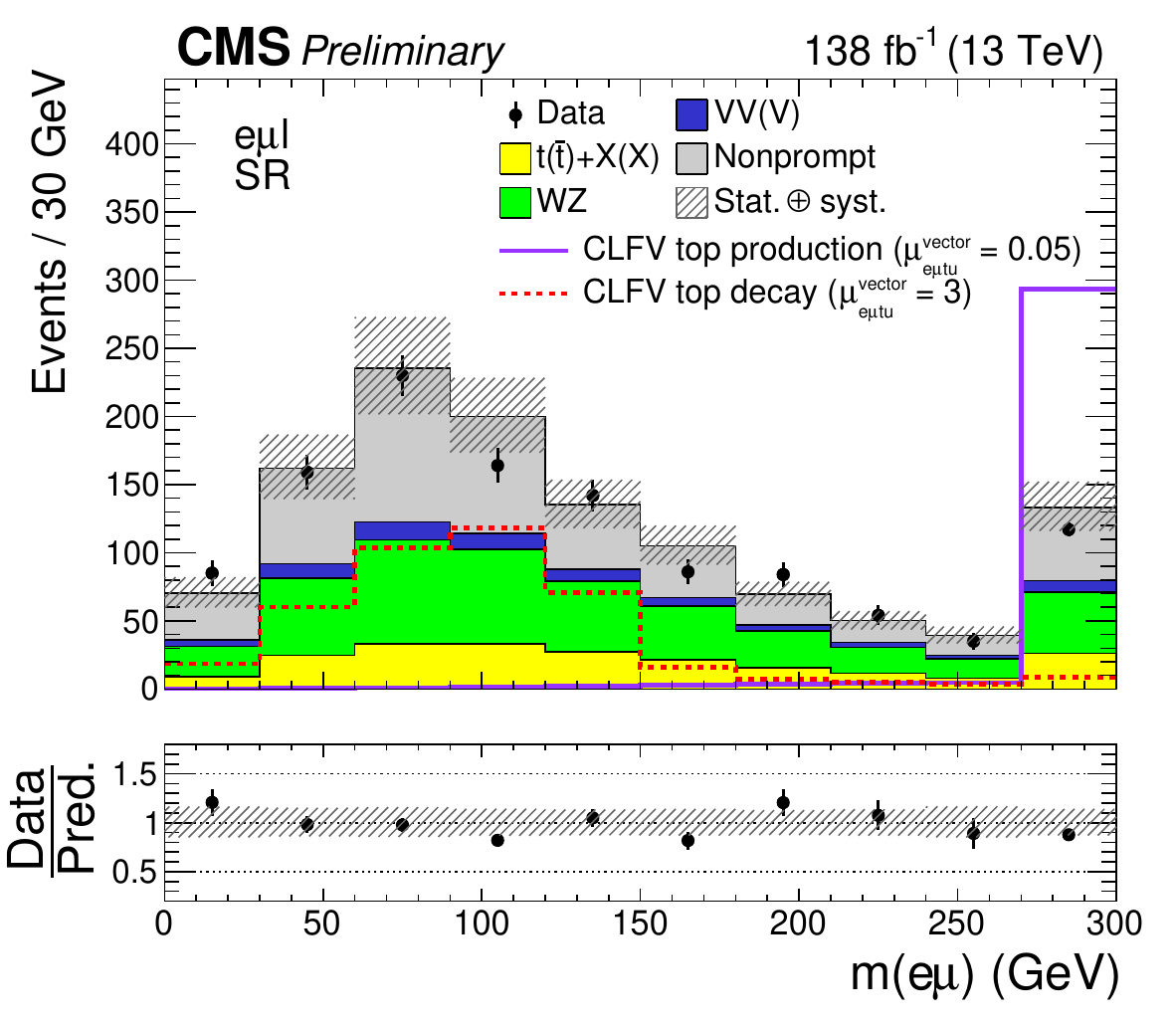}&
   \includegraphics[width=0.45\textwidth]{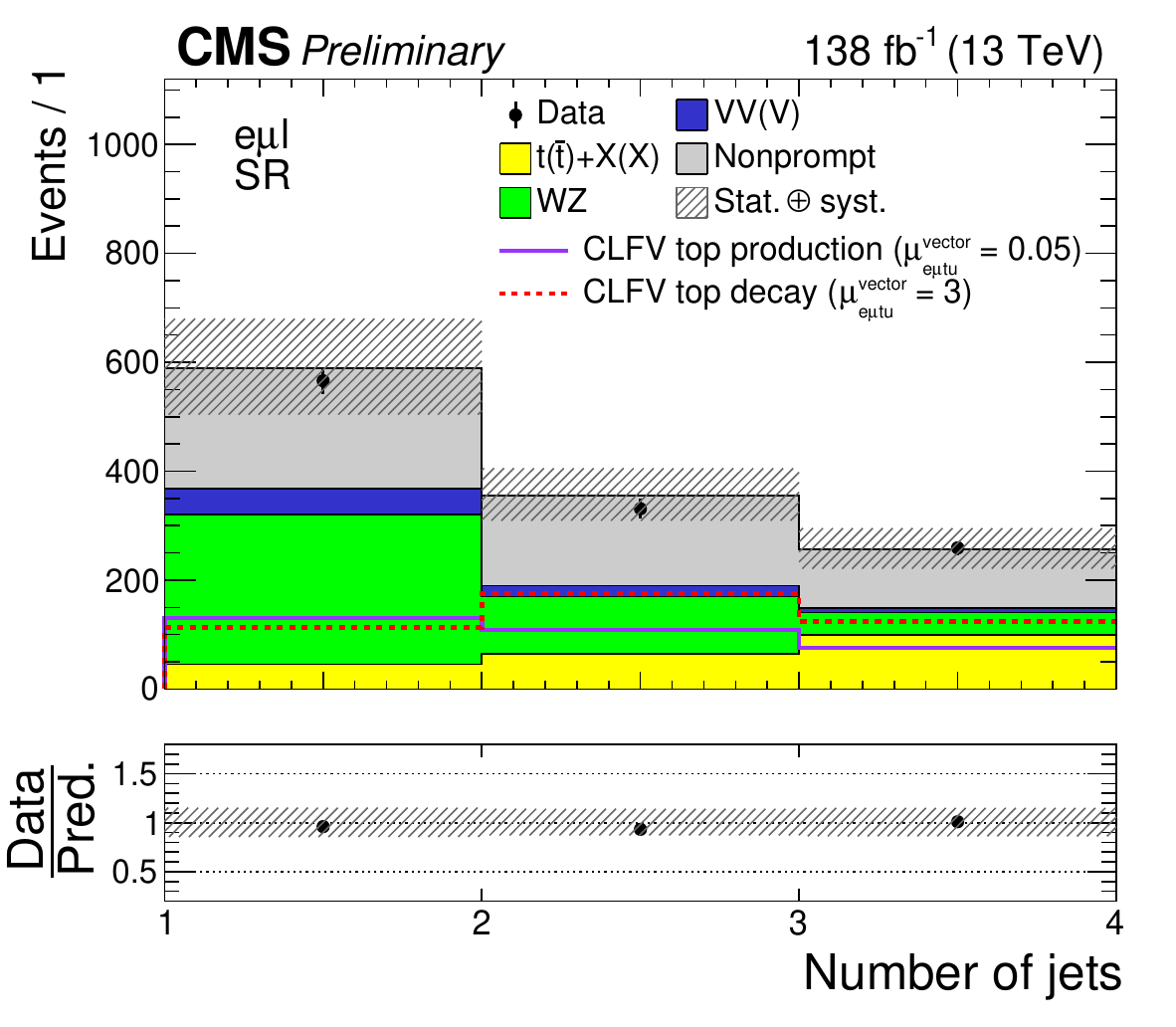}\\
  \end{tabular}
 \caption{Distributions of the CLFV $e\mu$ mass (left) and the number of jets (right) in the SR. The cross sections of the CLFV signals are scaled arbitrarily for an improved virtualization.}
 \label{fig:SR}
 \end{center}
\end{figure}

The kinematic distributions of the final-state particles differ significantly between two different signals. Most notable is the presence of high-$p_{T}$ leptons in the top production signals, which results in high $m(e\mu)$ (as is shown in Figure \ref{fig:SR}). On the contrary, the $m(e\mu)$ is bounded by the top quark mass ($m_{t}$) when the leptons come from CLFV top decay. Therefore, the SR is subdivided into two regions using a cutoff at m(e$\mu$) = 150 GeV, targeting two different signals.

\begin{figure}[h]
 \begin{center}
  \begin{tabular}{cc}
   \includegraphics[width=0.45\textwidth]{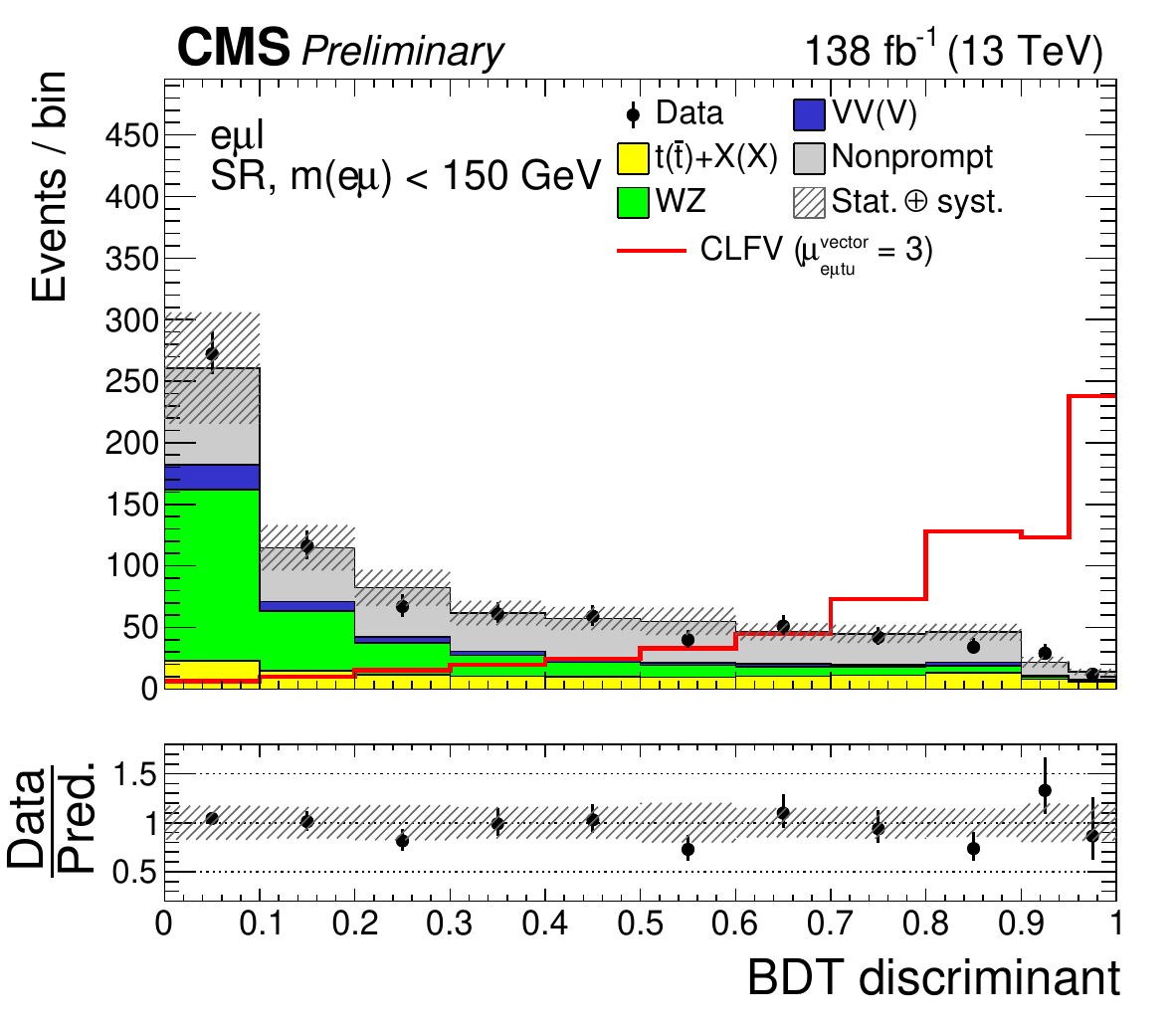}&
   \includegraphics[width=0.45\textwidth]{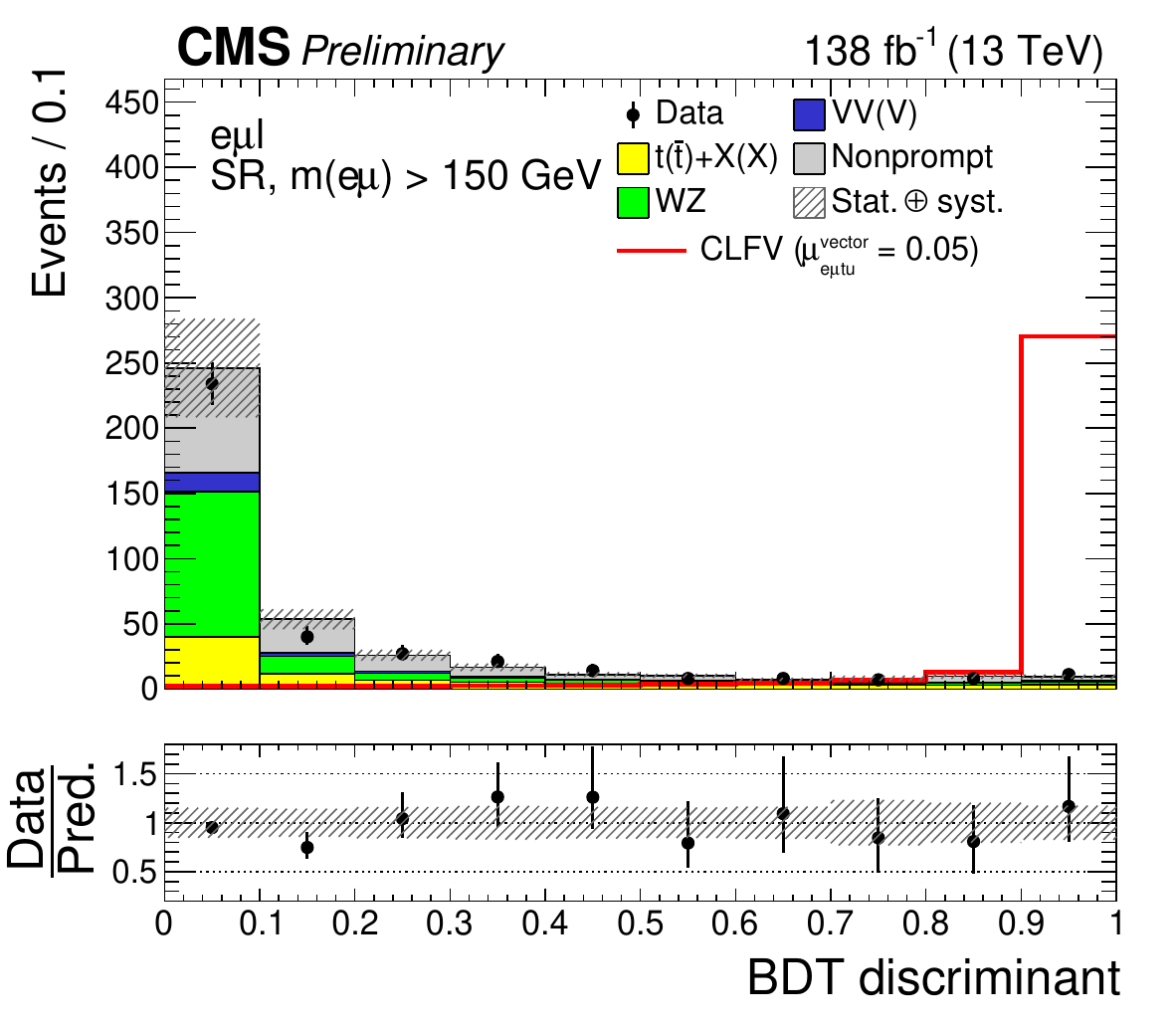}\\
  \end{tabular}
 \caption{Distributions of the BDT discriminator targeting the CLFV top quark decay (left) and production (right) signal. Contributions from the two signal modes are combined within each SR.}
 \label{fig:BDT}
 \end{center}
\end{figure}

To further separate a possible CLFV signal from the SM background contributions in the SR, a Boosted Decision Tree (BDT) was employed. A binary BDT is trained for each of the two SRs separately. The resulting output BDT distributions are shown in Figure \ref{fig:BDT}, with a good agreement between the data and background prediction.

A binned likelihood function $\mathcal{L}(\mu, \theta)$ is constructed to perform the statistical analysis on the BDT discriminator distributions. The compatibility between the data and the combined signal plus background expectation under the hypothesized value of the signal strength $\mu$ is quantified by a test statistic that considers the profile likelihood ratio $q(\mu)=-2\ln\mathcal{L}(\mu,\hat{\theta_{\mu}})/\mathcal{L}(\hat{\mu},\hat{\theta})$.

Using the asymptotic modified frequentist CL$_{s}$ method \cite{Stats} with the profile likelihood ratio as the test statistic, upper limits are placed on $\mu$ at the 95\% confidence level (CL). The one-dimensional upper limits on a given Wilson coefficient are obtained by taking the square root of the signal strength while setting other Wilson coefficients to zero. The upper limits on branching fractions is obtained assuming $m_{t}$=172.5 GeV, and an energy scale of $\Lambda$=1 TeV~\cite{Theory3}.

\begin{figure}[h]
 \begin{center}
 \begin{tabular}{cc}
  \includegraphics[width=0.48\textwidth]{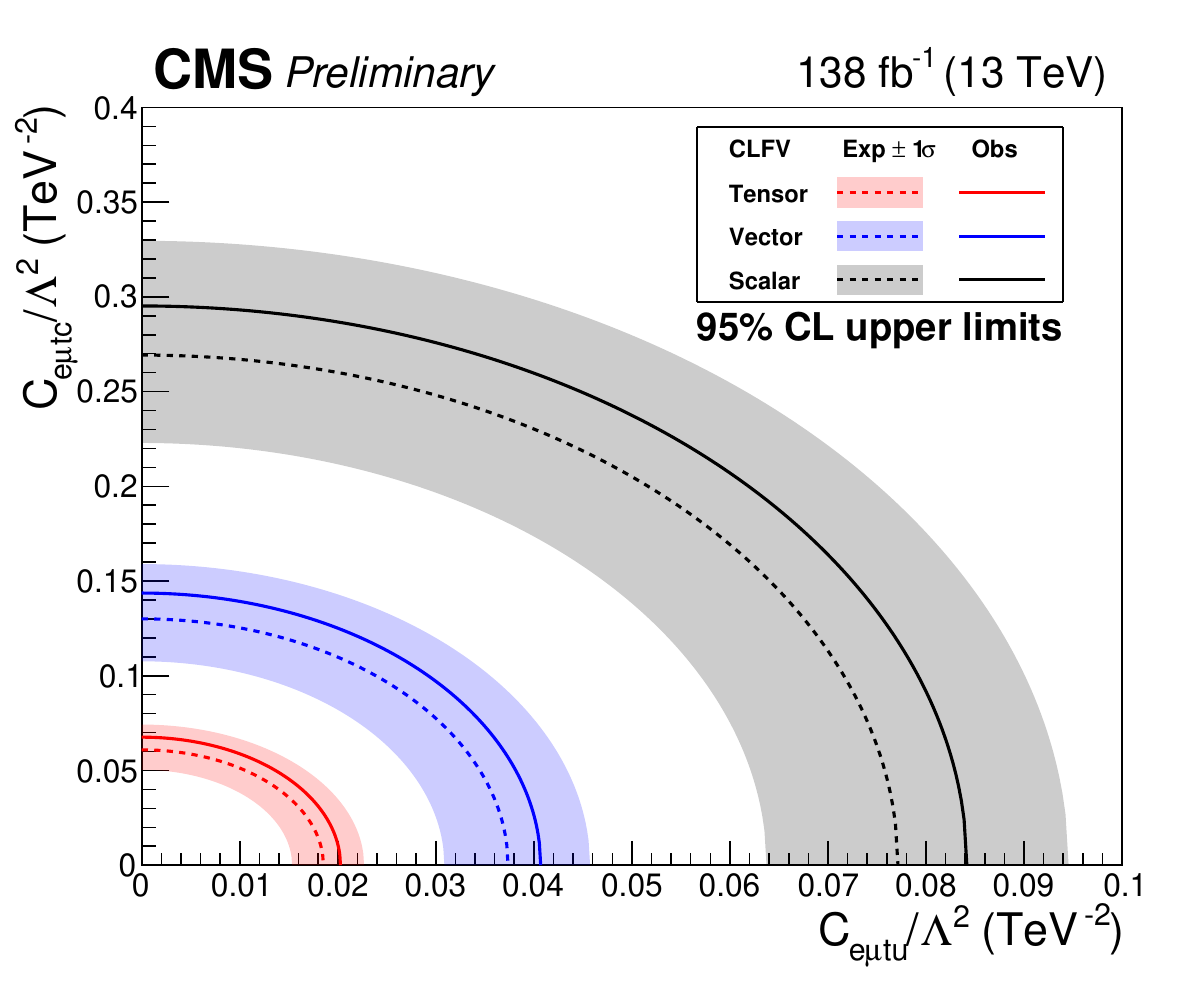}&
  \includegraphics[width=0.48\textwidth]{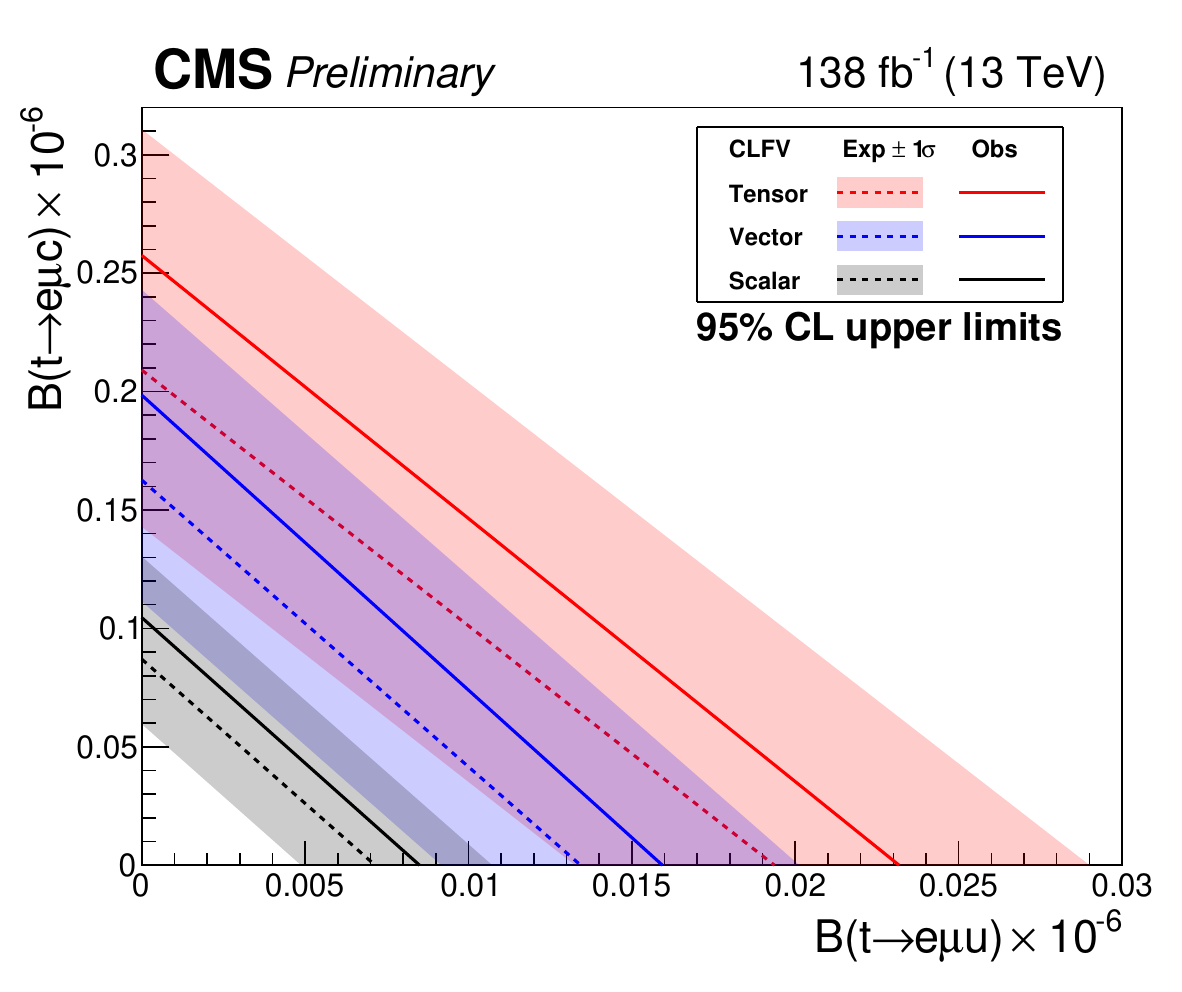}\\
 \end{tabular}
 \caption{Two-dimensional 95$\%$ CL upper limits on the Wilson coefficients (left) and the branching fractions (right). The shaded bands contain 68$\%$ of the distribution of the expected upper limits.}
 \label{fig:2dlimit}
 \end{center}
\end{figure}

The results for the one-dimensional limits are summarized in Table~\ref{tab:limit}. Assuming a linear relationship between $\mathcal{B}(t \rightarrow e\mu tu)$ and $\mathcal{B}(t \rightarrow e\mu tc)$ in the case of nonvanishing signals, the two-dimensional limits can also be obtained through interpolation (shown in Figure~\ref{fig:2dlimit}). 

\begin{table}[h]
\centering
\caption{Upper limits at 95\% CL on the different CLFV signals. The expected and observed upper limits are shown in regular and bold fonts, respectively.}
\begin{tabular}{cccccc}
CLFV       & Lorentz  & \multicolumn{2}{c}{$C_{e\mu tq}/\Lambda^2~(\mathrm{TeV}^{-2})$} & \multicolumn{2}{c}{$\mathcal{B} (t \rightarrow e\mu q) \times 10^{-6}$} \\
coupling  & structure & Exp (68\% range) & \textbf{Obs} & Exp (68\% range) & \textbf{Obs} \\
\noalign{\vskip 1mm}
\hline
\noalign{\vskip 1mm}
\multirow{3}{*}{$e\mu tu$}& tensor & 0.019 (0.015-0.023) & \textbf{0.020} & 0.019 (0.013-0.029) & \textbf{0.023}\\
& vector & 0.037 (0.031-0.046) & \textbf{0.041} & 0.013 (0.009-0.020) & \textbf{0.016}\\
& scalar & 0.077 (0.064-0.095) & \textbf{0.084} & 0.007 (0.005-0.011) & \textbf{0.009}\\
\noalign{\vskip 1mm}
\hline
\noalign{\vskip 1mm}
\multirow{3}{*}{$e\mu tc$} & tensor & 0.061 (0.050-0.074) & \textbf{0.068} & 0.209 (0.143-0.311) & \textbf{0.258}\\
 & vector & 0.130 (0.108-0.159) & \textbf{0.144} & 0.163 (0.111-0.243) & \textbf{0.199}\\
 & scalar & 0.269 (0.223-0.330) & \textbf{0.295} & 0.087 (0.060-0.130) & \textbf{0.105}\\
 \noalign{\vskip 1mm}
\hline
\end{tabular}
\label{tab:limit}
\end{table}

\section{Conclusion}

Preliminary results from a search for charged-lepton flavor violation are presented. An effective field theory approach is used for parametrizing the charged-lepton flavor violating interactions. A boosted decision tree is used to distinguish the signal from the background. No significant excess is observed over the expectations from the standard model. Upper limits are set on the branching fractions $t\rightarrow e\mu u$ ($t\rightarrow e\mu c$) of 0.023 $\times 10^{-6}$ (0.256$\times 10^{-6}$), 0.016 $\times 10^{-6}$ (0.199 $\times 10^{-6}$), 0.009 $\times 10^{-6}$ (0.105 $\times 10^{-6}$) for tensor, vector and scalar interactions, respectively.

%

\end{document}